\documentstyle[aaspp4]{article}

\received{}
\accepted{}

\slugcomment{to appear in ApJ 501, January 1999} 

\lefthead{Chang \& Ho}
\righthead{{\it RXTE} Observation of PSR B0656+14}

\newcommand{\beqary}{\begin{eqnarray}}
\newcommand{\eeqary}{\end{eqnarray}}
\newcommand{\beq}{\begin{equation}}
\newcommand{\eeq}{\end{equation}}

\newcommand{\cm}{{\rm cm}}

\newcommand{\s}{{\rm s}}

\newcommand{\keV}{{\rm keV}}

\newcommand{\fu}{\cm^{-2}\s^{-1}\keV^{-1}}

\begin{document}

\title{{\it RXTE} Observation of PSR B0656+14}

\author{Hsiang-Kuang Chang} 
\affil{Department of Physics, National
                 Tsing Hua University, Hsinchu 30043, Taiwan, ROC}
\and
\author{Cheng Ho}
\affil{Los Alamos National Laboratory, MS D436, Los Alamos, NM 87545, USA}

\begin{abstract}
PSR B0656+14 was observed by the {\it Rossi X-ray Timing Explorer (RXTE)} with 
the proportional counter array (PCA) and the high-energy x-ray timing experiment
(HEXTE) for 160 ksec during August 22 -- Septenber 3, 1997.
No pulsation was firmly found in the timing analysis, 
in which the contemporaneous radio ephemeris and various
statistical tests were applied  for searching evidence of pulsation. 
A marginal detection of pulsation at a confidence level of 95.5\%
based on the $H$-test was found with data in the whole HEXTE energy band.
In the energy band of 2-10 keV
the {\it RXTE} PCA upper limits are about one order of magnitude lower 
than that from {\it ASCA} GIS data.
If the {\it CGRO} EGRET detection of this pulsar is real,
considering the common trait that most EGRET-detected pulsars
have a cooling spectrum in hard x-ray and gamma ray energy bands,
the estimated {\it RXTE} upper limits indicate a deviation (low-energy 
turn-over) from 
a cooling spectrum starting from  20 keV or higher.
It in turn suggests an outer-magnetospheric synchrotron-radiation origin
for high-energy emissions from PSR B0656+14.
The {\it RXTE} PCA upper limits also suggest that a reported power-law component
based on {\it ASCA} SIS data in 1-10 keV fitted jointly with {\it ROSAT} data, 
if real, should be mainly unpulsed.
\end{abstract}


\section{Introduction}
\label{intro}

The pulsar PSR B0656+14
was discovered by Manchester et al.\ (1978),
with a period of 384.9 millisecond, a characteristic age of $1.1
\times 10^5$ yr and an inferred surface dipole magnetic field of $4.7
\times 10^{11}$ G.
After the report of its possible optical counterpart
(Caraveo, Bignami, \& Mereghetti 1994), its optical emission was recently
observed to be pulsed at the radio period and shows a nonthermal
origin (Shearer et al. 1997; Pavlov, Welty, \& C\'{o}rdova 1997;
Kurt et al. 1998).
X-ray pulsation at the period of PSR B0656+14 from the {\it EINSTEIN}
x-ray source E0656+14 was found by C\'{o}rdova et al.\ (1989).
Subsequent observations with {\it ROSAT} showed thermal characteristics 
with a hard tail in
its spectrum in 0.1-2.4 keV, which can be fit well by a 
helium-atmosphere blackbody model (Finley, \"{O}gelman, \& Kizilo\u{g}lu 1992),
or a two-component model comprising either two blackbodies 
or a blackbody plus a power law
(Possenti, Mereghetti, \& Colpi 1996).
Using {\it ACSA} data jointly with {\it ROSAT} data,
Greiveldinger et al.\ (1996) found that
a two-component model cannot fit the spectrum well,
and a three-component model (two blackbodies plus a power-law) 
gives an acceptable fit.
However, a recent study (Wang et al.\ 1998) did not confirm the necessity
of invoking the third (power-law) component.
At the high-energy end, PSR B0656+14 joined the EGRET-pulsar family with
 a relatively weak evidence of pulsation
(Ramanamurthy et al.\ 1996). Like other EGRET pulsars, it shows
a power-law spectrum in the EGRET energy band. 

With the proposed power-law component from fitting 
{\it ROSAT} + {\it ASCA} data
and that from the EGRET observation, it is apparent that the spectrum will
bend, either gradually or more abruptly, 
between 10 keV and 100 MeV (Figure~\ref{spectrum}).
Such a spectral bending can be used to  constrain possible emission sites
and mechanisms for these x-rays and gamma-rays (Chang \& Ho 1997). 
On the other hand, observations around 10 keV will help
to clarify the existence of the power-law component in the {\it ROSAT}
+ {\it ASCA} study.
In this {\it Letter} we report the results of a 160-ksec {\it RXTE} 
Cycle 2 observation of
PSR B0656+14 in the energy range of 2 - 250 keV.
Though no pulsation was detected,
our estimated upper limits for the pulsed flux are low enough to be used
to support 
an outer-magnetospheric synchrotron-radiation origin
for high-energy emissions from PSR B0656+14.
They also suggest that the reported power-law component
based on {\it ASCA} SIS data in 1-10 keV fitted jointly with 
{\it ROSAT} data, 
if real, should be mainly unpulsed.

\section{Observation}
\label{ob}  

The PCA and HEXTE on board {\it RXTE} were pointed at 
PSR B0656+14 during August
22 -- September 3, 1997 (MJD 50682 -- 50694), for about 160 ksec.
The {\it RXTE} mission, spacecraft and instrument
capabilities are described in Swank et al.\ (1995), Giles et al.\
(1995) and Zhang et al.\ (1993) The PCA consists of five essentially
identical PCUs with a total effective area of 6729 cm$^2$.
The HEXTE comprises two clusters of four scintillation counters.
The net open area of the eight detectors is 1600 cm$^2$
(Rothschild et al.\ 1998). The two instruments PCA and HEXTE have no
imaging capability. Their field of view is one degree.

Data were first screened according to the following two criteria:
(1) the offset, that is, the difference between the source position and the
pointing of the satellite, is less than 0.02 degree; and
(2) the elevation angle, that is, the angle between the Earth's limb
and the target viewed from the satellite, is larger than 10 degrees.
During the whole observation, the five PCUs of the PCA were all on.
The total exposure for the PCA is 152.4 ksec $\times$ 6729 cm$^2$, which
is equal to 1.026 $\times 10^9$ sec cm$^2$.
One detector in Cluster B of the HEXTE has lost its spectral capability,
and we excluded all the photons detected by that detector in our analysis.
The total exposure for the HEXTE is then 153.9 ksec $\times$ 1600 cm$^2
\times\frac{7}{8}$, which is 2.155 $\times 10^8$ sec cm$^2$. 
 
Around the same epoch of the {\it RXTE} observation, PSR B0656+14 was also
monitored at Jodrell Bank Radio Observatory. The radio ephemeris is
summarized in Table 1 and used as the input for pulsation search.
\placetable{ephemeris}

\section{Analysis and Results}
\label{ana}  

The data were reduced to the solar system barycenter and analyzed using the JPL
DE200 ephemeris, the pulsar position listed in Table~\ref{ephemeris},
and the {\it RXTE}-related tasks in the software package FTOOLS v.4.0.

In the timing analysis, PCA data were divided into five energy bands
with each covering
PCA channels 6-16 (2.1-6.1 keV), 17-26 (6.1-9.8 keV), 27-55 (9.8-20.5
keV), 56-107 (20.5-40.4 keV), and 108-249 (40.4-98.5 keV), respectively. 
The HEXTE data were also divided into three bands:
HEXTE channels 14-60 (15-60 keV),
61-121 (60-125 keV), and 122-234 (125-250 keV).  
To search for pulsation,
each detected photon was assigned an arrival phase $\phi$
based on the ephemeris in Table~\ref{ephemeris}, and 
$
\phi=\mbox{fractional part of}\,(\nu(t-t_0)+\dot{\nu}(t-t_0)^2/2
+\ddot{\nu}(t-t_0)^3/6)
$,
with $\nu$ being the pulsar frequency, $t$ the barycentric corrected 
arrival time of the
photon, and $t_0$ the radio epoch.
We then binned the photons in each energy band and
various combinations of energy bands  according to
 their arrival phases to form lightcurves
with 20 and 40 bins in one period.
The result was that none of the lightcurves
shows significant deviation from a model steady distribution under
the Pearson's $\chi^2$-test (Leahy et al.\ 1983a,b). 

We also applied the
bin-independent parameter-free $H$-test (De Jager, Swanepoel \&
Raubenheimer 1989) to search for signature of pulsation. 
The $H$-test was applied to the data in different energy bands and
various combinations.  The results of the $H$-test all show high
probability of the data being consistent with a steady source.
Only a very marginal detection of pulsation was found in the whole 
HEXTE 
energy band (channels 14-234, 15-250 keV) with a 
95.5\% probability of being
inconsistent with a steady source based on the $H$-test. 
The folded lightcurve in this energy band is shown in Figure~\ref{lightcurve}.
Applying the straight $Z_1^2$-test
(the Rayleigh test, which is more appropriate if the underlying pulse
profile is sinusoidal), the probability is 
90.6\%, while the $Z_3^2$-test gives a probability of 98.4\%. 
\placefigure{lightcurve}

In our analysis,
the $\chi^2$-test and the $H$-test were repeated in an interval of
 pulsation frequencies
near the radio one.
Unlike the spiky dependence of the $\chi^2$-value on frequencies,
the $H$-test gives a smooth variation in the corresponding probability for
different frequencies. 
None of these trials gives significant evidence of pulsation.
A higher probability was found at a frequency lower than the radio one
by $9\times 10^{-8}$ Hz. At this frequency, the $H$-test gives a
98.2\% probability of pulsation detection in the HEXTE data covering
channels 14-234.

Based on these analyses, we do not consider the current observation
provides evidence of
pulsation from PSR B0656+14 in the {\it RXTE} energy band.

The upper limit of pulsed flux is estimated following the prescription
given by Ulmer et al. (1991)
Assuming a duty cycle of 0.5 we 
obtain the following 3-$\sigma$ upper limits, which are also shown
in Figure~\ref{spectrum}: 
for PCA, $1.3\times 10^{-6}\,\fu$ in 2.1-6.1 keV,
$1.1\times 10^{-6}\,\fu$ in 6.1-9.8 keV, $5.6\times
10^{-7}\,\fu$ in 9.8-20.5 keV, $3.8\times 10^{-7}\,\fu$ in
20.5-40.4 keV, and $1.4\times 10^{-7}\,\fu$ in 40.4-98.5 keV;
for HEXTE, 
$1.1\times 10^{-6}\,\fu$ in 15-60 keV,
$6.6\times 10^{-7}\,\fu$ in 60-125 keV, and $3.4\times
10^{-7}\,\fu$ in 125-250 keV. 
\placefigure{spectrum}

\section{Discussion}
\label{dis}  

In the study with {\it ROSAT} PSPC and {\it ASCA} SIS data,
Greiveldinger et al.\ (1996) reported
a three-component spectral model, which consists of two blackbodies with
temperature at $7.8\times 10^5$ K and $1.5\times 10^6$ K and a power-law
with a photon spectral index of -1.5.  
Based on the {\it ASCA} GIS data, the pulsed fraction was estimated
to be 71\% as a 3-$\sigma$ upper limit in 0.5-10 keV, 
or 31$\pm$10\% at a 2-$\sigma$ level of detection in 0.5-2 keV and
100\% as a 3-$\sigma$ upper limit in 2-10 keV.
For comparison, the proposed power-law component, which includes both
pulsed and unpulsed fluxes, and the 3-$\sigma$ upper
limit in 2-10 keV from {\it ASCA} GIS data for the pulsed flux are plotted 
in Figure~\ref{spectrum}.
Our estimated 3-$\sigma$ upper limits from {\it RXTE} PCA data 
in 2-10 keV are about one order of magnitude lower than that from
 {\it ASCA} GIS data. 
These new upper limits make it plausible that the spectrum of the pulsed flux 
below 20 keV, if detected, should have a photon spectral index larger than
-1.5, assuming that the index increases from -2.8 above 100 MeV 
monotonically toward the lower-energy end.    

EGRET pulsars
all have
power-law spectra typically covering two orders of magnitude in the EGRET
band with best-fit photon spectral indices in the range of $-1.4$ to
$-1.8$ except for the Crab and PSR B0656+14 (Fierro 1995; Nolan et al.\ 1996;
Merck et al.\ 1996). 
Their spectra all turn flatter at lower energies.
For example, PSR B1951+32, though with a steeper spectrum in the EGRET band, 
has a spectrum with a photon spectral index of -1.5 around 1-10 MeV.
This spectral behavior has been discussed and used to form an argument which
provides a useful diagnostic to constrain possible emission locations and
mechanisms for these high-energy emissions 
(Chang \& Ho 1997).
The key issue in that argument is a `bending energy' below which the spectrum
starts deviating from a cooling one.
A cooling spectrum is due to a cooling model of
a steady-state electron/positron distribution and has a photon spectral index of 
$-3/2$ for the dominant cooling mechanism being synchrotron radiation 
 and $-5/3$ for
curvature radiation. 
In the parameter space of the electron/positron
energy and the distance from the stellar center, which characterizes the
strength of magnetic fields and the curvature radius of fieldlines,
one can find a line along which
the radiative cooling time scale for a certain radiation mechanism 
is comparable to the dynamical time scale
of the relativistic motion of the charges. 
The observed radiation is expected to occur around a distance 
charaterized by   
a point on that equal-time-scale 
line at which the critical energy of radiated photons for the corresponding
parameters is equal to the observed bending energy. 
At any other distances, the cooling population of radiating charges,
if exists, will have a cooling spectrum with a bending energy either higher
or lower than the observed one.
In view of the common trait in the spectral behavior of most EGRET
pulsars, it is very likely that if PSR B0656+14 has a spectrum with a photon
spectral index around -1.5
at some energies below 100 MeV, it must start getting
flatter toward low energy from above 20 keV as required by 
the {\it RXTE} PCA upper limits.
A bending energy higher than 20 keV for PSR B0656+14 suggests
an emission location at a distance larger than a couple of 
$10^8$ cm from the stellar
center and the synchrotron radiation being the dominant emission mechanism
for radiating charges with a typical pitch angle of about 0.1 to 0.001.  
Curvature radiation does not yield a reasonable emission location within
the light cylinder. 
More observations with the {\it RXTE} and the future {\it INTEGRAL}
 are very much desired for providing
better statistics and helping advance our understanding of
high-energy emissions from neutron stars. 

Up to date, existing observations  show that except for the Crab,
the bending energies for other gamma-ray pulsars are all higher than
a few tens of keV, though many of them are not yet well determined.
On the other hand, some observations also reveal possible nonthermal x-ray
emissions around keV from some of the gamma-ray pulsars.
However, they may have  an origin different from that of the whole continuum
from several tens of keV up to a few GeV.
For example, the reported nonthermal emissions around 1 keV
for the Geminga and PSR B1951+32   
both have a photon spectral index of about -1.5 (Halpern \& Wang 1997; 
Safi-Harb, \"{O}gelman \& Finley 1995), which,
together with their flux magnitudes,
makes them a separate component
from that of higher energies; see Figure 6 in Strickman et al. (1996) and
Figure 2 in Chang \& Ho (1997). 

The current observation, since no pulsation was detected,
cannot confirm the power-law component proposed 
in Greiveldinger et al. (1996) for PSR B0656+14.
However, the estimated upper limits ensure that if that component is
real, it must be mainly unpulsed.
An unpulsed power-law spectrum is not easy to be attributed to
magnetospheric emissions.  
Though in the analysis of Wang et al. (1998)
 the power-law component was not confirmed,
its existence and being unpulsed could just be a good example supporting 
the idea
of the `electron-positron blanket' proposed by Wang et al. (1998)
Future {\it AXAF} and {\it XMM} observations
will be able to solve this issue.



\acknowledgments
We thank Andrew Lyne for
providing
the radio ephemeris.
This
work was performed under the auspices of the US Department of Energy
and was supported in part by the {\it RXTE} Guest Observer Program. 
HKC is supported by the NSC of ROC under the grant NSC-88-2112-M-007-050. 

\clearpage

\figcaption[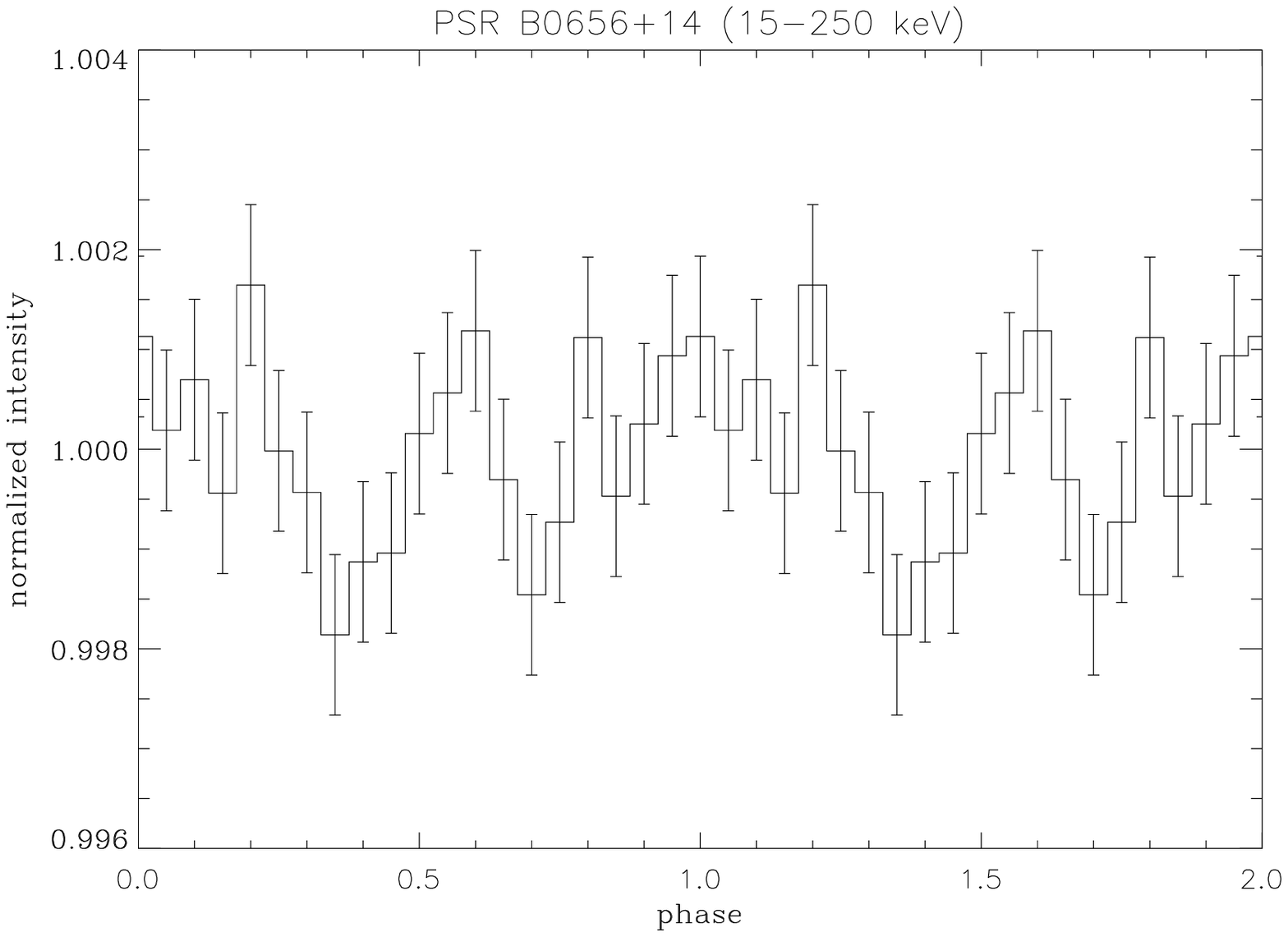]{
The folded lightcurve of PSR B0656+14 from {\it RXTE} HEXTE data.
\label{lightcurve}
}

\figcaption[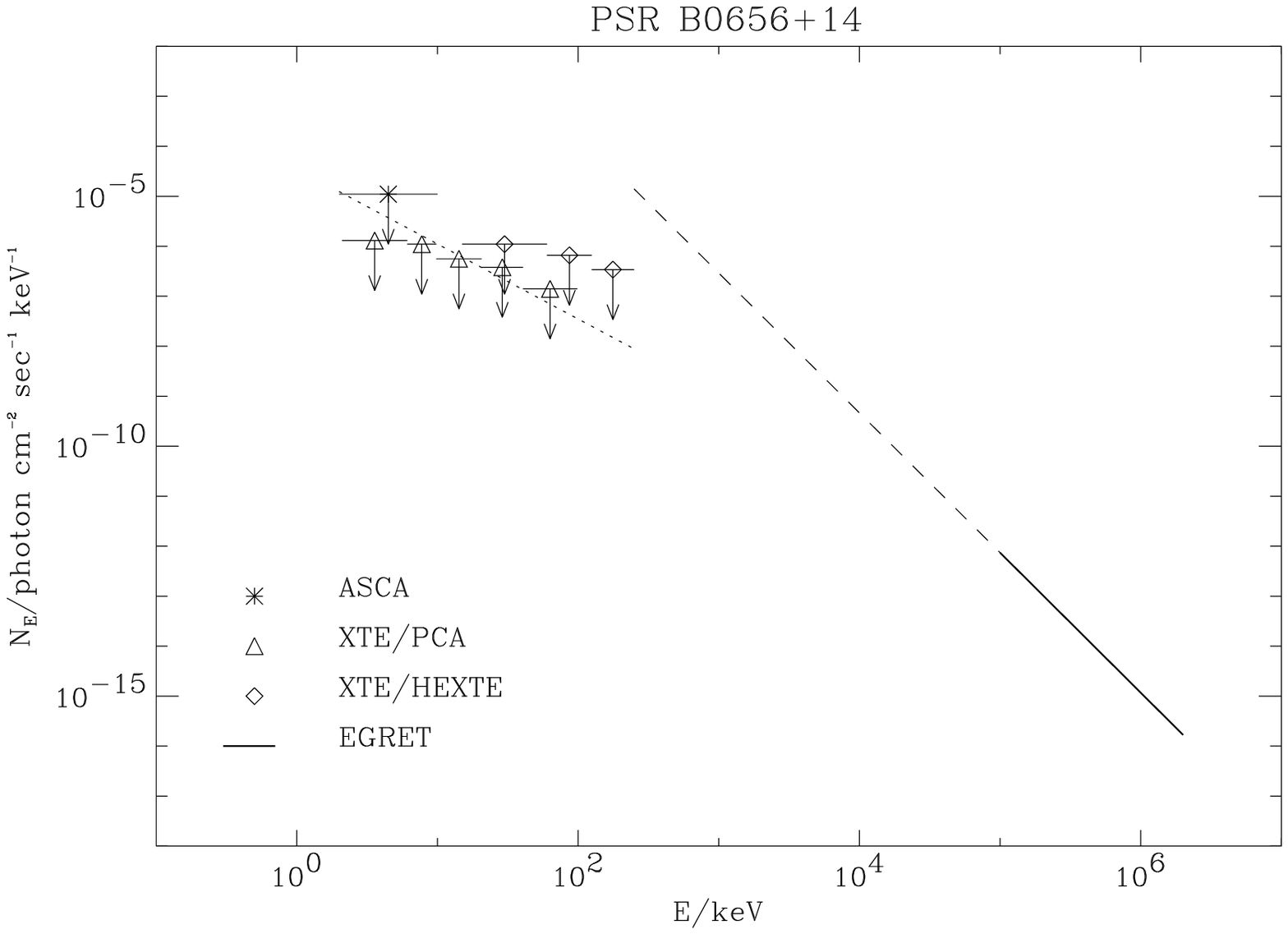]{
The spectrum of PSR B0656+14 from 2 keV up to 2 GeV.
Upper limits are all for pulsed flux and at a 3-$\sigma$ level. 
The {\it ASCA} upper limit in 2-10 keV is based on the {\it ASCA} GIS data
(Greiveldinger et al.\ 1996). 
The dotted line is the power-law component with a photon spectral index of -1.5
proposed by
Greiveldinger et al.\ (1996) from fitting {\it ASCA} SIS data in 
1-10 keV together with {\it ROSAT} PSPC data,
which includes both pulsed and unpulsed components.
It was not confirmed by a later analysis (Wang et al.\ 1998).
This power law is plotted in this figure with an extension to a higher
energy.
It intersects with the EGRET spectrum at about 100 MeV.
\label{spectrum}
}

\clearpage

\begin{deluxetable}{llr}
\tablecaption{
Radio ephemeris of PSR B0656+14\tablenotemark{a}
\label{ephemeris}}
\tablehead{
\colhead{} & \colhead{} & \colhead{} 
} 
\startdata
Validity interval & (MJD) & 50606 - 50797 \nl
Epoch, $t_0$ & (MJD) & 50701.000002341 \nl
$\alpha_{2000}$ & & $6^{\rm h}59^{\rm m}48^{\s}.126$ \nl
$\delta_{2000}$ & & $14^{\circ}14'21''.15$ \nl
$\nu$ & (Hz) & 2.5981054226747 \nl
$\dot{\nu}$ & (Hz/s) & $-3.71150\times 10^{-13}$ \nl
$\ddot{\nu}$ & (Hz/s/s) & $8.33\times 10^{-25}$ \nl 
\enddata
\tablenotetext{a}{provided by Andrew Lyne (1998, private communication)}
\end{deluxetable}

\clearpage
\plotone{lightcurve.eps}
\clearpage
\plotone{spectrum.eps}

\begin{thebibliography}{}

\bibitem[1999]{xyz}
 Buccheri, R., et al.\ 1983, A\&A, 128, 245
\bibitem[1999]{xyz}
 Caraveo, P. A., Bignami, G. F., \& Mereghetti, S. 1994, ApJ, 422, L87
\bibitem[1999]{xyz}
 Chang, H.-K., \& Ho, C. 1997, ApJ, 479, L125
\bibitem[1999]{xyz}
 C\'{o}rdova, F. A., et al.\ 1989, ApJ, 345, 451
\bibitem[1999]{xyz}
 De Jager, O.C., Swanepoel, J.W.H., \& Raubenheimer, B.C. 1989, A\&A, 221, 180
\bibitem[1999]{xyz}
 Fierro, J.M. 1995, Ph.D. thesis, Stanford University
\bibitem[1999]{xyz}
 Finley, J. P., \"{O}gelman, H., \& Kizilo\u{g}lu, \"{U}. 1992, ApJ, 394, L21
\bibitem[1999]{xyz}
 Giles, A.B., Jahoda, K., Swank, J.H., \& Zhang, W. 1995,
 Publ.\ Astron.\ Soc.\ Australia, 12, 219
\bibitem[1999]{xyz}
 Greiveldinger, C., et al.\ 1996, ApJ, 465, L35
\bibitem[1999]{xyz}
 Halpern, J. P., \& Wang, F. Y.-H. 1997, ApJ, 477, 905
\bibitem[1999]{xyz}
 Kurt, V. G., Sokolov, V. V., Zharikov, S. V., Pavlov, G. G., \&
 Komberg, B. V. 1998, A\&A, 333, 547
\bibitem[1999]{xyz}
 Leahy, D.A., et al.\ 1983a, ApJ, 266, 160
\bibitem[1999]{xyz}
 Leahy, D.A., Elsner, R.F., \& Weisskopf, M.C. 1983b, ApJ, 272, 256
\bibitem[1999]{xyz}
 Manchester, R. N., et al.\ 1978, MNRAS, 185, 409
\bibitem[1999]{xyz}
 Merck, M., et al.\ 1996, A\&AS 120, C465
\bibitem[1999]{xyz}
 Nolan, P. L., et al.\ 1996, A\&AS 120, C61
\bibitem[1999]{xyz}
 Pavlov, G. G., Welty, A. D., \& C\'{o}rdova, F. A. 1997, ApJ, 489, L75
\bibitem[1999]{xyz}
 Possenti, A., Mereghetti, S., \& Colpi, M. 1996, A\&A, 313, 565
\bibitem[1999]{xyz}
 Ramanamurthy, P. V., Fichtel, C. E., Kniffen, D. A., Sreekumar, P.,
 \& Thompson, D. J. 1996, ApJ, 458, 755
\bibitem[1999]{xyz}
 Rothschild, R. E., et al.\ 1998, ApJ, 496, 538
\bibitem[1999]{xyz}
 Safi-Harb, S., \"{O}gelman, H., \& Finley J. P. 1995, ApJ, 439, 722
\bibitem[1999]{xyz}
 Shearer, A., et al. 1997, ApJ, 487, L181
\bibitem[1999]{xyz}
 Strickman, M. S., et al.\ 1996, ApJ, 460, 735
\bibitem[1999]{xyz}
 Swank, J.H., Jahoda, K., Zhang, W., \& Giles, A.B. 1995,
 in The Lives of the Neutron Stars, ed.\ M.A. Alpar, \"U. Kizilo\u{g}lu,
 \& J. van Paradijs (NATO ASI Series C, 450)(Boston: Kluwer), 525
\bibitem[1999]{xyz}
 Ulmer, M.P., Purcell, W.R., Wheaton, W.A., \& Mahoney, W.A. 1991, ApJ, 369, 485
\bibitem[1999]{xyz}
 Wang, F. Y.-H., Ruderman, M., Halpern, J. P., \& Zhu, T. 1998, ApJ, 498, 373
\bibitem[1999]{xyz}
 Zhang, W., et al.\ 1993, Proc.\ SPIE, 2006, 324 

\end{thebibliography}
\end{document}